\documentclass{ctr_summer}

\usepackage{ctrfont}
\usepackage{natbib}
\usepackage{undertilde}
\usepackage{color}

\usepackage[dvips]{graphicx}
\usepackage{psfrag}
\usepackage{epsfig}
\usepackage{amsmath,rotating}
\usepackage{dashrule}
\usepackage{undertilde}

\usepackage{url}

\usepackage{color}

\newcommand{\vqow}{\ensuremath{\vec{q}_{\text{ow}}}}
\newcommand{\vq}{\ensuremath{\vec{q}}}
\newcommand{\ua}{\ensuremath{\langle u \rangle}}
\newcommand{\usa}{\ensuremath{\langle u^2 \rangle}}

\title{Toward real-time high-fidelity simulation using integral
boundary layer modeling}

\shorttitle{Integral boundary layer wall modeling}

\author{A.\ Marques, Q.\ Wang\footnote[1]{Department of Aeronautics and Astronautics, MIT}, J.\ Larsson\footnote[2]{Department of Mechanical Engineering, University of Maryland}, G.\ Laskowski\footnote[3]{GE Aviation} \and S.\ Bose\footnote[4]{Cascade Technologies Inc.}}

\shortauthor{Marques et~al.}

\begin{document}

\setcounter{page}{1}

\maketitle

One of the greatest challenges to using large-eddy
simulations (LES) in engineering applications is the large
number of grid points required near walls. To mitigate this
issue, researchers often couple LES with a simplified model
of the flow close to the wall, known as the wall model. One
feature common to most wall models is that the first few
(about 3) grid points must be located below the inviscid
log-layer ($y/\delta  < 0.2$), and the grid must have
near-isotropic resolution near the wall. Hence, wall-modeled
LES may still require a large number of grid points in both
the wall-normal and span-wise directions. Because of these
requirements, wall-modeled LES is still unfeasible in many
applications. We present a new formulation of wall-modeled
LES that is being developed to address this issue. In this
formulation, LES is used to solve only for the features of
the velocity field that can be adequately represented on the
LES grid. The effects of the unresolved features are
captured by imposing an integral balance of kinetic energy
in the near-wall region. This integral energy balance
translates into a dynamic partial differential equation
defined on the walls, which is coupled to the LES equations.
We discuss details of the new formulation and present
results obtained in laminar channel flows.\\

\hrule

\section{Introduction}\label{sec:intro}
One of the greatest challenges to using of large-eddy
simulations (LES) in engineering applications is the large
number of grid points required to resolve the small scales
related to viscous effects near solid walls. To mitigate
this issue, researchers often couple LES with a simplified
model of the flow close to the wall, known as the wall
model~\citep{piomelli:2002, sagaut:2006, kawai:2010}. One
feature common to most wall models is that the first few
(about 3) grid points must be located below the inviscid
log-layer ($y/\delta < 0.2$), and the grid must have
near-isotropic resolution near the wall. Hence, wall-modeled
LES may require a large number of grid points in both the
wall-normal and span-wise directions. Because of these
requirements, wall-modeled LES is still unfeasible in many
applications.

We are developing a new formulation of wall-modeled,
time-filtered LES to address this issue. While the
formulation is aimed at turbulent flows, so far the
development has been restricted to laminar channel flows. In
certain configurations, the dynamic evolution of laminar
channel flows results in very large velocity gradients.
These large gradients are due to viscous effects restricted
to the near-wall region, and depend mostly on the localized
features of the flow, which are amenable models that are
relatively independent of specific applications. These are
the same kinds of models that we intend to develop for
turbulent flows. Furthermore, the relative simplicity of
laminar channel flows allows us to create and test models
with low computational cost in this initial stage of
development.

The new wall model formulation is based on the decomposition
of the velocity field (time-filtered in the turbulent case)
into two components
\begin{equation}\label{eq:decomposition}
 \vq = \vqow + \vq_{\text{nwd}}.
\end{equation}
The off-wall component ($\vqow$) includes only the large
flow scales prevalent away from the wall, whereas the
near-wall defect component $\vq_{\text{nwd}}$ takes into
account the small near-wall scales not included in the
off-wall component. In performing this decomposition we
operate under the assumption that, at least away from solid
walls, the long time scales are associated with large
spatial scales, and we use a finite volume discretization
on a relatively coarse grid to compute the off-wall
component. Note that there is no grid resolution
error incurred in this step, since the off-wall component
includes only scales that can be resolved by the coarse
grid. 

We supplement the governing equations of the off-wall
component with data-driven models of the effects of the
unresolved small scales near the wall. We build these models
based on relationships measured on numerical data obtained
in fully resolved simulations. Furthermore, these models are
parametrized by a low-dimensional feature space that depends
only on localized integral flow quantities. Hence, we solve
only for the integral quantities of the near-wall defect
component that affect the governing equations of the
off-wall component. As discussed in
Section~\ref{sec:formulation}, in the case of laminar
channel flows it is sufficient to augment the system with
the integral balance of kinetic energy in the first grid
cell above the wall. This approach is conceptually similar
to the classical integral boundary layer formulation used to
model viscous effects coupled with inviscid flow
formulations~\citep{drela:bl:1987}.

The remainder of this report is organized as follows. In
Section~\ref{sec:heat} we discuss the laminar channel flows
that are the subject of the current investigation. We
present in detail the new wall model formulation in
Section~\ref{sec:formulation} and show preliminary results
in Section~\ref{sec:results}. In Section~\ref{sec:steps} we
list the steps needed to advance the development of the
formulation, and in Section~\ref{sec:conclusions} we make
final remarks.

\section{Laminar channel flow}\label{sec:heat}

The present investigation was primarily based on laminar
channel flows. The combination of periodicity and lack of
turbulence reduces the formulation of such flows to a simple
one-dimensional heat equation, given by
\begin{equation}\label{eq:heat}
 \dfrac{\partial u}{\partial t} 
 - \nu\dfrac{\partial^2 u}{\partial y^2} =
 -\dfrac{1}{\rho}\dfrac{dp}{dx},
\end{equation}
where $dp/dx$ denotes the constant pressure gradient that
drives the flow. Despite the simplicity of
Eq.~(\ref{eq:heat}), this model problem includes the
viscous dissipation effects that dominate the dynamics in
the near-wall region.

In certain circumstances, Eq.~(\ref{eq:heat}) can lead to
the development of viscous boundary layers with very large
velocity gradients. In principle, resolving these large
gradients requires very fine grids. The objective of the
present investigation is to create a wall model formulation
that allows the solution of such flows with relatively
coarse grids, such that these gradients are under-resolved
or not resolved at all. Furthermore, we want to develop
a formulation that can be extended to more complex physics,
including turbulent flows.

\begin{figure}
 \begin{center}
 \includegraphics[width=0.6\textwidth]{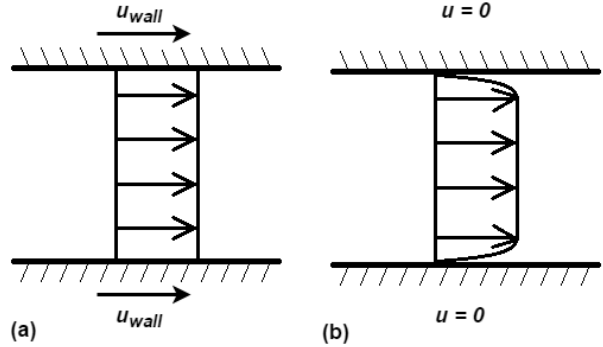}
 \caption{Prototype laminar channel flow problem with large
          velocity gradient. (a) Flow is initially uniform
          moving along with wall. (b) Immediately after
          the wall stops large velocity gradients develop
          in the near-wall region.}
 \label{fig:heat}
 \end{center}
\end{figure}

One situation in which Eq.~(\ref{eq:heat}) can lead to large
velocity gradients is illustrated in Figure~\ref{fig:heat}.
This figure depicts the situation in which, initially, the
channel walls are moving with a constant speed
$u_{\text{wall}}$, and the channel flow is uniform with the
same speed as the walls. In this setting, the flow is
completely driven by the motion of the walls, and there is
no pressure gradient along the channel. Then, the walls are
brought to a sudden stop. In the moments that follow the
stopping of the walls, the flow immediately adjacent to the
wall is forced to stop due to friction, whereas the flow at
the bulk of the channel continues to move at a speed close
to $u_{\text{wall}}$. This situation results in very large
velocity gradients in the wall-normal direction. We use this
prototype flow to assess the performance of our wall model
formulation.

\section{Formulation}
\label{sec:formulation}

In practice, the velocity decomposition in
Eq.~(\ref{eq:decomposition}) is obtained by defining the
off-wall velocity component. The near-wall defect component
follows from the difference between the off-wall and the
(unknown) complete velocity fields. Our experience showed
that an effective way to define the off-wall velocity
component is to apply a spatial filter to the complete
velocity field. By judiciously selecting the spatial filter,
we can always guarantee that the off-wall component can be
adequately represented in a relatively coarse grid.

As mentioned above, the present investigation is restricted
to laminar channel flows. However, in the case of turbulent
flows we expect the effect of the spatial filter on the
already time-filtered velocity to be restricted to the
near-wall region. This follows from the assumption that,
away from the wall, the long time scales are related to
large spatial scales. We will examine the validity of this
assumption in future work.

In this investigation we adopted the grid cell averaging as
the spatial filter, a common approach in LES
formulations~\citep{moin:1991, sagaut:2006, you:2007}. The
filtered velocity at the $i$-th cell is given by
\begin{equation}\label{eq:average}
 \ua_i = \dfrac{1}{V_i}\int_{V_i} u\,dV,
\end{equation}
where $V_i$ denotes the volume of the $i$-th cell. When
applied to a fine grid, the filtered velocity satisfies the
standard finite volume discretization of the governing
equations. On the other hand, when applied to a coarse grid,
the fluxes between grid cells in the near-wall region depend
on small-scale features that are not resolved in the
off-wall velocity component. Hence, standard approximations
used to compute the fluxes between grid cells are not
accurate in the near-wall region, and we must supplement the
finite volume discretization with more accurate flux
estimates. Furthermore, these estimates must be based on
relationships between these small-scale features and the
large-scale features resolved in the coarse grid. These
relationships can be measured in fully resolved simulations.

In the case of laminar channel flows, the finite volume
discretization of Eq.~(\ref{eq:heat}) results in
\begin{equation}\label{eq:fv}
 \dfrac{d \ua_i}{dt}
 - \dfrac{F_{i+1/2} - F_{i-1/2}}{\Delta y_i} =
 - \dfrac{1}{\rho}\dfrac{dp}{dx},
\end{equation}
where $\Delta y_i$ is the length of the $i$-th grid cell and
$F$ denotes the momentum flux between grid cells. For grid
cells away from the wall, $F$ is approximated by
\begin{equation}\label{eq:F}
 F_{i+1/2} = \dfrac{2\nu(\ua_{i+1} - \ua_i)}
                   {\Delta y_{i+1} + \Delta y_i}.
\end{equation}
In the current investigation we supplement the finite volume
discretization with models for the skin friction coefficient
and the momentum flux between the first and second cells,
given by
\begin{align}
 C_{f_0} &= \dfrac{\nu}{0.5 u_r^2}
        \left.\dfrac{\partial u}{\partial y}\right|_{y=0},
 \label{eq:cf0}\\
 C_{f_1} &= \dfrac{\nu}{0.5 u_r^2}
            \left.\dfrac{\partial u}{\partial y}
            \right|_{y=\Delta y_1},
 \label{eq:cf1}
\end{align}
where
\begin{equation}\label{eq:ur}
 u_r = \sqrt{\ua_1 + \ua_2}
\end{equation}
is a reference velocity scale. We parametrized these models
using an approach analogous to the one used in the classical
integral boundary layer formulation~\citep{drela:bl:1987}.
Namely, we define quantities analogous to the displacement
($\delta^{\ast}$) and momentum thicknesses ($\theta$) as
follows
\begin{align}
 \delta^{\ast} &=
 \left(1-\dfrac{\ua_1}{u_r}\right)\Delta y_1,
 \label{eq:delta}\\
 \theta &=
 \left( \dfrac{\ua_1}{u_r} 
      - \dfrac{\usa_1}{u_r^2}\right)\Delta y_1.
  \label{eq:theta}
\end{align}
We then normalize the quantities of interest with respect
to $\theta$ and $u_r$ and measure the following
relationships
\begin{align}\label{eq:model}
 Re_{\theta}C_{f_0} &= Re_{\theta}C_{f_0}(H),&
 Re_{\theta}C_{f_1} &= Re_{\theta}C_{f_1}(H),
\end{align}
where $H = \delta^{\ast}/\theta$ is the shape factor, and
$Re_{\theta} = u_r\theta/\nu$ denotes the Reynolds
number with respect to $\theta$.

\begin{figure}
 \begin{center}
 \includegraphics[width=0.9\textwidth]{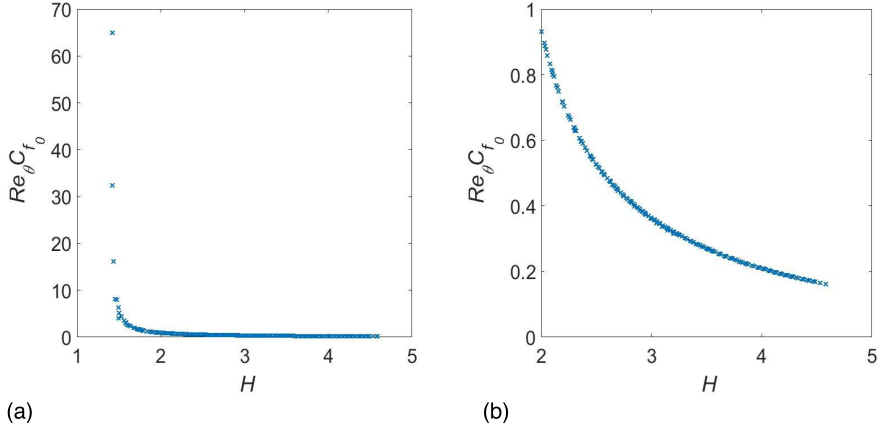}
 \caption{Skin friction coefficient, normalized by $u_r$ and
          $\theta$, as a function of the shape factor $H$.
          (a) Full range of $H$.
          (b) Enlarged view of the range $H > 2$.}
 \label{fig:cf0}
 \end{center}
\end{figure}

\begin{figure}
 \begin{center}
 \includegraphics[width=0.45\textwidth]{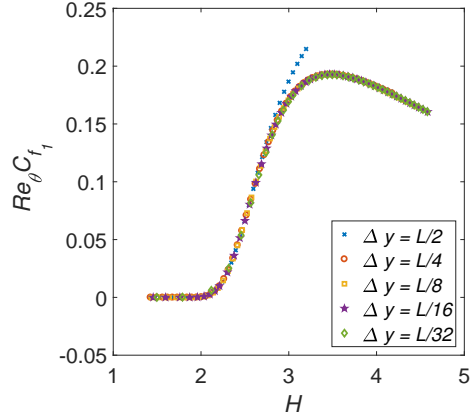}
 \caption{Momentum flux between first and second grid cells,
          normalized by $u_r$ and $\theta$ as a function of
          the shape factor $H$. Results for several values
          of the wall-normal grid spacing, as a fraction of
          the channel width $L$.}
 \label{fig:cf1}
 \end{center}
\end{figure}

Figures~\ref{fig:cf0} and \ref{fig:cf1} show the values of
$Re_{\theta}C_{f_0}$ and $Re_{\theta}C_{f_1}$ as a
function of the shape factor. These values are based on a
solution of Eq.~(\ref{eq:heat}) with a fine grid with 1,025
nodes. Furthermore, $C_{f_1}$ is measured between the first
and second grid cells of a coarse grid. Figure~\ref{fig:cf1}
shows $Re_{\theta}C_{f_1}$ for several values of coarse grid
spacing. Nevertheless, after the normalization all curves
nearly collapse. The exception is $\Delta y = 0.5L$ at higher
values of $H$.

Equations~(\ref{eq:fv})-(\ref{eq:model}) do not result in a
closed system. Namely, Eq.~(\ref{eq:theta}) depends on
$\usa_1$, related to the cell average of kinetic energy in
the first grid cell, and is not readily available as a
solution of Eq.~(\ref{eq:fv}). We solve for $\usa_1$ by
augmenting Eq.~(\ref{eq:fv}) with an integral balance of
kinetic energy at the first grid cell. Namely, the evolution
$\usa_1$ is obtained by multiplying Eq.~(\ref{eq:heat}) by
$u$ and computing the average over the first grid cell, as
follows
\begin{equation}\label{eq:ke}
 \dfrac{d\usa_1}{dt} - u_r^3 C_D =
 -\dfrac{1}{\rho}\dfrac{dp}{dx}\ua_1,
\end{equation}
where
\begin{equation}
 C_D = \dfrac{\nu}{0.5 u_r^3}
       \int_0^{\Delta y_1}
        u\dfrac{\partial^2 u}{\partial y^2}\,dy
\end{equation}
is the viscous dissipation coefficient. We can also rewrite
Eqs.~(\ref{eq:fv})-(\ref{eq:ke}) in terms of the evolution
of the shape factor $H$ as follows,
\begin{multline}\label{eq:dHdt}
 \dfrac{1}{Re_{\theta}}\dfrac{dH}{dt}
 - \dfrac{2H C_D}{\theta^2} 
 + 2H(C_{f_1} - C_{f_0})
   \left[ \dfrac{2}{\theta^2} - \dfrac{2H}{\Delta y_1\theta} 
        + \dfrac{H(H+1)}{\Delta y_1^2} \right]\\
 - 2\dfrac{\ua_2}{u_r}(\bar{C}_{f_2} - C_{f_1})
   \left[\dfrac{H(H+1)\theta + \Delta y_1(1-H)}
               {\Delta y_2 \theta^2}\right]\\
 = -\dfrac{\nu}{\theta\rho u_r^2}\dfrac{dp}{dx}
   \left[ H(H+1)\left( \dfrac{\ua_2}{u_r} 
                     - \dfrac{H^2\theta}{\Delta y_1} \right)
        + (1-H)\dfrac{\ua_2}{u_r}\dfrac{\Delta y_1}{\theta}
    \right],
\end{multline}
where
\begin{equation}
 \bar{C}_{f_2} = \dfrac{\nu F_{2.5}}{0.5 u_r^2}.
\end{equation}

\begin{figure}
 \begin{center}
 \includegraphics[width=0.9\textwidth]{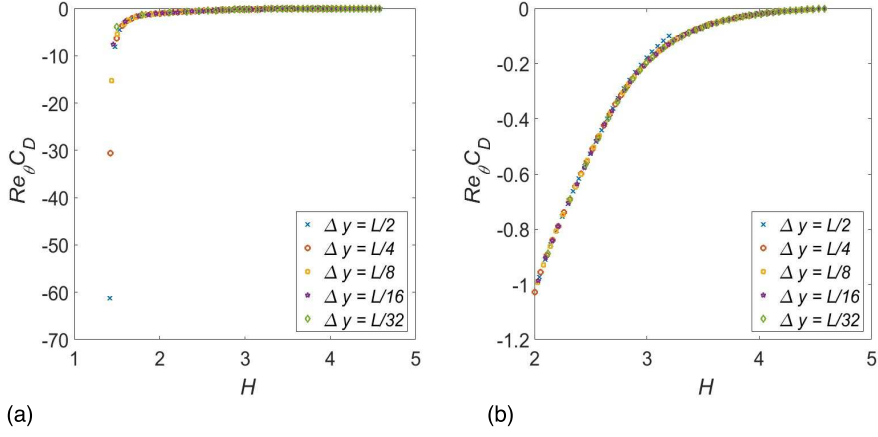}
 \caption{Viscous dissipation in the first grid cell,
          parametrized by the shape factor $H$.
          (a)~Full range of $H$.
          (b) Enlarged view of range $H > 2$.}
 \label{fig:cd}
 \end{center}
\end{figure}

Finally, the viscous dissipation coefficient $C_D$ also
needs to be modeled as a function of the shape factor, after
proper normalization. Figure~\ref{fig:cd} shows
$Re_{\theta}C_D$ as a function of the shape factor for
several values of coarse grid spacing. This figure shows
that, after appropriate normalization, the results for $C_D$
also nearly collapse into a single curve.

The formulation described above can be interpreted as a
coupled system of governing equations for the off-wall
velocity component and an integral feature that depends on
the near-wall defect (cell-averaged kinetic energy). The
near-wall defect component affects the off-wall equations
in two ways: (i) a Neumann boundary condition in the form
of a prescribed shear stress at the wall, and
(ii) additional volumetric source terms that model the
momentum flux between neighboring grid cells in the
near-wall region (here we restricted the effect of this
source term to the first two grid cells). On the other hand,
the off-wall component affects the governing equation of the
near-wall component in the form of source terms in
Eq.~(\ref{eq:dHdt}).

\section{Results}\label{sec:results}

\begin{figure}
 \begin{center}
 \includegraphics[width=0.9\textwidth]{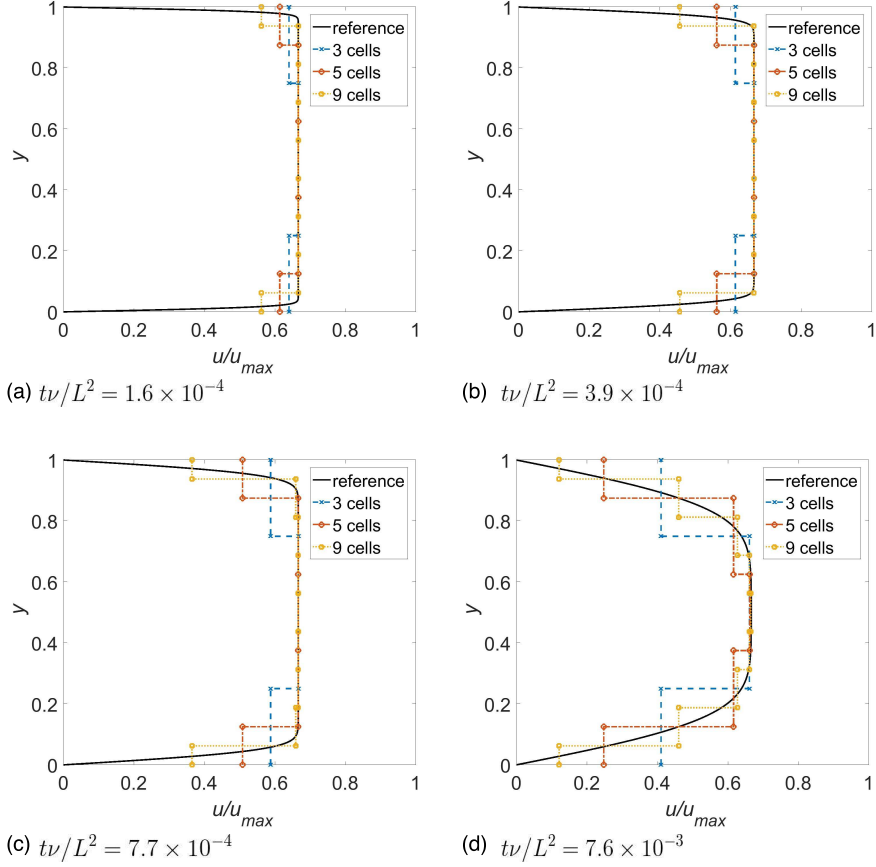}
 \caption{Solution of a laminar channel flow brought to a
          sudden stop. Comparison between a refined solution
          (1,025 nodes) and solutions obtained with the new
          wall model at
          (a)~$t\nu/L^2 = 1.6 \times 10^{-4}$,
          (b)~$t\nu/L^2 = 3.9 \times 10^{-4}$,
          (c)~$t\nu/L^2 = 7.7 \times 10^{-4}$, and
          (d)~$t\nu/L^2 = 7.6 \times 10^{-3}$.}
 \label{fig:snapshots}
 \end{center}
\end{figure}

\begin{figure}
 \begin{center}
 \includegraphics[width=0.9\textwidth]{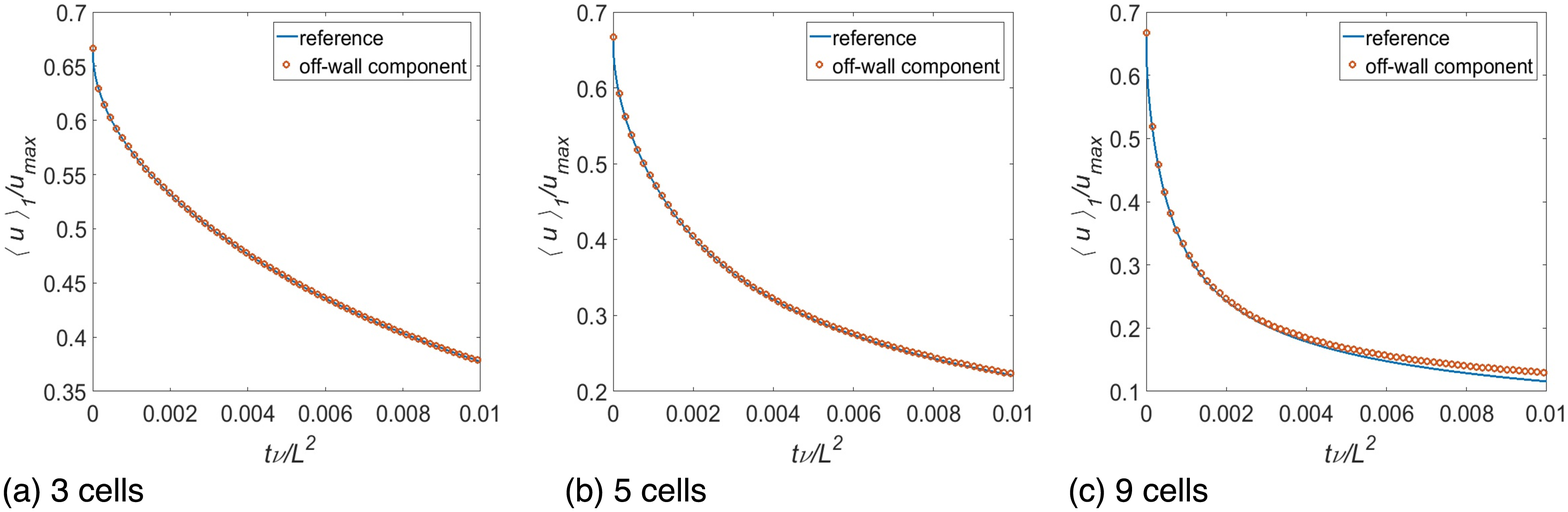}
 \caption{Cell average velocity at the first grid cell.
          Comparison between solutions obtained with the
          new wall model formulation with coarse grids
          -- (a) 3, (b) 5, and (c) 9 grid cells --  and
          values computed directly from a reference solution
          obtained with 1,025 grid nodes.}
 \label{fig:v0}
 \end{center}
\end{figure}

\begin{figure}
 \begin{center}
 \includegraphics[width=0.9\textwidth]{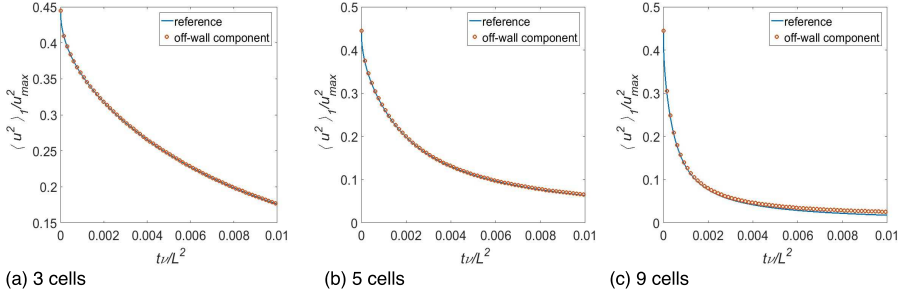}
 \caption{Cell average kinetic energy at the first grid cell.
          Comparison between solutions obtained with the
          new wall model formulation with coarse grids
          -- (a) 3, (b) 5, and (c) 9 grid cells --  and
          values computed directly from a reference solution
          obtained with 1,025 grid nodes.}
 \label{fig:w0}
 \end{center}
\end{figure}

We verified the validity of the formulation discussed in
Section~\ref{sec:formulation} in a series of test cases
involving laminar channel flows in which the walls are
brought to a sudden stop, as described in
Section~\ref{sec:heat}. Figure~\ref{fig:snapshots} shows
snapshots of the off-wall velocity component at different
moments in time. We can visually inspect that the off-wall
velocity component is close to the reference solution
averaged over the grid cells. Furthermore,
Figures~\ref{fig:v0} and \ref{fig:w0} show comparisons of
$\ua_1$ and $\usa_1$ computed with the new formulation and
values computed directly from the reference solution. We can
observe that the current formulation produces accurate
results in the initial development of the boundary layer,
when large velocity gradients are observed.

\section{Next steps}\label{sec:steps}

The formulation presented in Section~\ref{sec:formulation}
proved successful to compute solutions of laminar channel
flows even when large velocity gradients are grossly
under-resolved. However, preliminary results (not shown
here) indicate that simulations may present instabilities if
time integration is carried out long enough. It is paramount
that we determine whether this behavior stems from numerical
instabilities that can be remedied, or from accrual of
errors from the models used to supplement the cell-averaged
equations.

Furthermore, so far we restricted our attention to the model
problem of laminar channel flows. However, the goal of this
research is to develop effective wall models for simulating
turbulent flows. Hence, the natural extension of this
work is to apply an analysis similar to the one presented in
Section~\ref{sec:formulation} to channel flows in the
turbulent regime. The challenge in this development lies in
the fact that, contrary to the simple model problem studied
in this work, the near-wall dynamics of turbulent flows are
not restricted to one-dimensional viscous effects. Hence,
additional terms are needed to supplement the cell-averaged
governing equations. In addition, these terms may
potentially depend on a higher-dimensional feature space
that has yet to be identified.

Finally, due to the chaotic nature of turbulent flows, we
anticipate that relationships such as the ones shown in
Figures~\ref{fig:cf0}-\ref{fig:cd} will not collapse so
tightly as observed in the laminar regime. In this situation,
we foresee the application of supervised learning techniques
to create the models needed to supplement the cell-averaged
governing equations. This approach can create models based
on the large amounts of DNS data available for turbulent
channel flows~\citep{perlman:2007, li:2008, graham:2013},
and results in stochastic models that allows us to quantify
the model inadequacy related to the use of wall
models in turbulent simulations. A similar technique has
been applied in the context of the classical integral
boundary layer formulation~\citep{marques:2016}.

\section{Conclusions}\label{sec:conclusions}

We developed a new wall model formulation for laminar
channel flows that produced accurate results even when large
velocity gradients are grossly under-resolved. This work
constitutes the initial stage in the development of wall
models for LES simulations of turbulent flows that are
nearly independent of the Reynolds number.

The formulation developed here is based on a decomposition
of the velocity field into (i) the cell average (off-wall
component) and (ii) the defect between the cell average and
the complete velocity (near-wall defect component). The
cell average quantities are computed with a finite volume
discretization of the governing equations, supplemented with
models for the shear stress and viscous dissipation in the
near-wall region. These models represent the effect of the
defect component on the cell averages, and work effectively
as boundary conditions and additional source terms applied
to the finite volume discretization.

Furthermore, these models for shear stress and viscous
dissipation in the near-wall region are created based on
numerical data obtained from fully resolved simulations.
In the case of laminar channel flows, it was possible to
parametrize this information with a single parameter, which
is analogous to the shape factor defined in classical
boundary layer theory.

We are currently working to extend the formulation discussed
here to solve turbulent flows in situations in which
geometrically thin boundary layers are under-resolved or
not resolved at all. We believe that the combination of a
formal velocity decomposition and data-driven models of the
defect terms has the potential to produce accurate wall
models in such situations, and results in a wall-modeled LES
formulation that meets the cost and accuracy demands of
practical engineering applications.

\subsection*{Acknowledgments}
The authors are grateful to the many participants of the CTR summer
program who engaged in thoughtful discussions about this project,
especially Prof.\ Peter Schmid.

\end{document}